\magnification=\magstep1
\baselineskip=18truept
\input amssym.def
\input psfig
\font\bfmathit=cmmib10

\tolerance=10000
\def\bibliography{\begingroup\baselineskip=15truept
        {}
        \vskip12truept}
\def\ref#1#2{\indent\llap{#1\enspace}\hang#2}
\def\endbibliography{\vfill\eject\endgroup}
\def\cL{{\cal L}}
\def\div{{\rm div}}
\def\sp{{^\prime}}
\def\spp{^{\prime\prime}}
\def\bR{{\bf R}}
\def\ra{{\rangle}}
\def\la{{\langle}}
\def\Uav{\hbox{$U_{\rm av}$}}
\centerline {\bf A connection between the Camassa-Holm equations and}
\medskip
\centerline {\bf turbulent flows in channels and pipes}
\vskip30pt
\centerline {S. Chen\footnote{$^1$}{Theoretical Division and Center
for Nonlinear Studies, Los
Alamos National Laboratory, Los Alamos, NM 87545},
C. Foias$^{1,}$\footnote{$^2$}{Department of Mathematics, Indiana
University, Bloomington, IN 47405},
D.D. Holm$^{1}$,
E. Olson$^{1,2}$,
E.S. Titi\footnote{$^{3}$}{%
Departments of Mathematics, Mechanical and Aerospace Engineering,
University of California, Irvine, CA 92697.}%
$^,$\footnote{$^4$}{Institute for
Geophysics and Planetary Physics, Los Alamos National Laboratory, Los
Alamos, NM 87545},
S. Wynne$^{3,4}$}
\bigskip\centerline{
Revised version, March 6, 1999.}

\vskip30pt

\noindent\underbar{Abstract}.  In this paper we discuss recent progress in
using
the Camassa-Holm equations to model turbulent flows.  The Camassa-Holm
equations,
given their special geometric and physical properties, appear
particularly well suited for studying turbulent flows.  We identify
the steady solution of the Camassa-Holm equation with the mean flow
of the Reynolds equation and compare the results with empirical data
for turbulent flows in channels and pipes.  The data suggests that the constant
$\alpha$ version of the Camassa-Holm equations, derived under the assumptions
that the fluctuation statistics are isotropic and homogeneous, holds to order
$\alpha$ distance from the boundaries. Near a boundary, these assumptions
are no
longer valid and the length scale $\alpha$ is seen to depend on the distance to
the nearest wall. Thus, a turbulent flow is divided into two regions:  the
constant $\alpha$ region away from boundaries, and the near wall region.
In the
near wall region, Reynolds number scaling conditions imply that $\alpha$
decreases as Reynolds number increases.  Away from boundaries, these
scaling conditions imply $\alpha$ is independent of Reynolds number.
Given the agreement with empirical and numerical data, our current
work indicates that the Camassa-Holm equations provide a promising
theoretical framework from which to understand some turbulent flows.

\bigskip


\vfill\eject

\noindent \underbar{Introduction}.  Laminar Poiseuille flow
occurs when a fluid in a straight channel, or
pipe, is driven by a constant upstream pressure gradient, yielding a
symmetric parabolic streamwise velocity profile.  In turbulent
states, the mean streamwise velocity remains symmetric, but is
flattened in the center because of the increase of the velocity
fluctuation.  Although a lot of research has been carried out for
turbulent channel flow,$^{1-6}$ accurate measurement of the mean
velocity and the Reynolds stress profiles, in particular for flows at
high Reynolds numbers, is still an experimental challenge.  However,
in the case of pipe flow, recent experiments for measuring the mean
velocity profile have been successfully performed for moderate to
high Reynolds numbers by Zagarola$^{7}$  The fundamental understanding of how
these profiles change as functions of Reynolds number, however, seems to be
still
missing.

In wall bounded flows it is customary to define a characteristic
velocity $u_*$ and wall-stress Reynolds number $R_0$ by $u_* =
\sqrt{|\tau_0|/\rho}$ and $R_0 = du_*/\nu$, where $\tau_0$ is the
boundary shear stress. We take the density $\rho$ to be unity, $\nu$
is the molecular viscosity of the fluid, and $d$ is a characteristic
macrolength.  For instance, for channel flow $d$ is the channel
half-width, and for pipe flow $d$ is the pipe radius.  Based on
experimental observation and numerical simulation, a piecewise
expression of the mean velocity across the channel or the pipe
has been commonly accepted,$^{8}$ for which the nondimensional
mean streamwise velocity,
$\phi \equiv U/u_*$, is assumed to depend on $\eta \equiv u_*z/\nu$
and have three types of behavior depending on the distance away from
the wall boundary, $z\;$:  a viscous sublayer, in which $\phi \sim
\eta$; the von K\'arm\'an-Prandtl logarithmic ``law of the wall,'' in
which $\phi(\eta) = \kappa^{-1}{\rm ln}\eta + A$, where $\kappa  \simeq$
0.41 and $A\simeq$ 5.5; and a power law region, in which $\phi \sim
\eta^p, 0 < p < 1$.  Alternatively, a single curve fitting over the whole
region
may be proposed (see Panton$^{9}$).  Yet another possibility is a family of
power
laws that fits the data away from the viscous sublayer, and has the log law
as an
envelope, as proposed by Barenblatt et al.$^{10}$

In this paper (a summary of which was given earlier$^{11}$), we
propose the viscous Camassa-Holm equations (VCHE) in (2.14) as a closure
approximation for the Reynolds equations.  The analytic form of our
profiles based on the steady VCHE away from the viscous sublayer, but
covering at least 95\% of the channel, depends on two free parameters:
the flux Reynolds number $R = d \Uav/\nu$ (where $\Uav$ is the
streamwise velocity, averaged across the channel), and the wall-stress Reynolds
number $R_0$.  Due to measurement limitations most experimental data are
contained
in this region.  Let us remark that we can further reduce the parameter
dependence to one free parameter by using a drag law for the wall
friction $D \sim R^2_0/R^2$.  For the remaining part of the channel, we
are unable to solve explicitly for the mean profile without further
assumptions, but we do show compatibility of the steady VCHE
with empirical and numerical velocity profiles in this subregion.
The VCHE profiles agree well with data obtained from measurements and
simulations of turbulent channel and pipe flow.
For another global approach to turbulent flows in channels
and pipes displaying good agreement of theoretical mean velocity profiles with
experimental data see Markus and Smith.$^{12}$

\vskip18pt

\noindent 1.  \underbar{The Euler-Poincar\'e equations and the Euler
equations}.  Consider the Lagrangian comprised of fluid kinetic energy and the
volume preservation constraint
$$\eqalignno{
L &= \int da \left\{{1\over 2} \left|{d\over dt} X(t,a)\right|^2 +
q(X(t,a),t) (\det X^\prime_a (t,a) - 1)\right\} \cr
&= \int dx \left\{{D\over 2} |u(x,t)|^2 + q(x,t)(1- D(x,t))\right\}\;.
&(1.1)\cr}$$
In (1.1),
$X(t,a)$ is the Lagrangian trajectory of the
fluid parcel starting at position $a$ at time $t=0$. Other notation is
$$\eqalignno{
X^\prime_a &= \nabla_aX,
\qquad
u(x,t) = {d\over dt}X(t,a)\;\cr
{\rm and}\;\;
D(x,t) &= (\det X^\prime_a (t,a))^{-1}\;\; {\rm at}\;\; x = X(t,a)\;.
&(1.2)\cr}$$
Moreover, the Jacobian $D$ satisfies the equation
$${\partial\over\partial t}D + \nabla\cdot(Du) = 0\;. \eqno(1.3)$$
The extremality conditions for $u$, where $q$ is viewed as a Lagrange
multiplier, are given by the Euler-Poincar\'e equation [13]
$$\left({\partial\over\partial t} + (u\cdot \nabla)\right) {1\over D}
{\delta\cL \over \delta u} + {1\over D} {\delta\cL \over \delta u_j}
\nabla u_j - \nabla {\delta\cL \over \delta D} = 0\;, \eqno(1.4)$$
(above and throughout we use Einstein's
notation for summations) and
$${\delta\cL \over \delta q} = 0\;. \eqno(1.5)$$
Since
$${1\over D} {\delta\cL\over \delta u} = u\;,\quad {\delta\cL \over
\delta D} = {1\over 2}\,u\cdot u-q\;,\quad {\delta\cL \over \delta q} =1
-D\,,$$
the relations (1.3), (1.4), (1.5) yield the Euler equations
$$\left({\partial\over\partial t} + u\cdot \nabla\right)u = -\nabla
q, \nabla\cdot u = 0 \;.$$
{\bf Remark.} The Euler-Poincar\'e equation (1.4) is equivalent in the Eulerian
picture to the corresponding Euler-Lagrange equation for fluid parcel
trajectories for Lagrangians such as (1.1) that are invariant under the
right-action of the diffeomorphism group, see Holm et al.$^{13-15}$
and references therein. In what follows, we shall introduce random fluctuations
into the description of the fluid parcel trajectories in the Lagrangian L in
(1.1), take its statistical average and use the Euler-Poincar\'e equation (1.1)
to derive Eulerian closure equations for the corresponding averaged fluid
motions.

\vskip18pt

\noindent 2.  \underbar{Averaged Lagrangians and the Camassa-Holm
equations}.  In the presence of random fluctuations the Lagrangian trajectory
given by $X(t,a)$ has to be augmented with fluctuations as
$$X^\sigma (t,a) = X(t,a) + \sigma (X(t,a),t)\;. \eqno(2.1)$$
Here $\sigma  = \sigma (x,t) = \sigma (x,t;\omega )$ is a random vector
field.  Thus the Lagrangian $L = L(\omega )$ becomes a random variable
$$L(\omega ) = \int da\left\{{1\over 2} \left|{d\over dt} X^\sigma
(t,a)\right|^2 + q^\sigma (X^\sigma  (t,a),t) [\det(X^\sigma
)^\prime_a (t,a) - 1]\right\}\;. \eqno(2.2)$$
In (2.2), we introduce the Eulerian velocity field,
$$u^\sigma (y,t) = {d\over dt} X^\sigma (t,a)\;\;{\rm for}\;\; y =
X^\sigma (t,a)\;, \eqno(2.3)$$
with $X^\sigma (t,a)$ given in equation (2.1).
This is similar to the classical Reynolds decomposition of fluid
velocity into its mean and fluctuating parts. However, this decomposition is
applied on Lagrangian fluid parcels, rather than at fixed Eulerian spatial
positions.

Introducing the decomposition (2.3) into the Lagrangian $L$ in (2.2) and
changing
the variables $a$ to $x = X(t,a)$ yields
$$L(\omega ) = \int dx \left\{{D\over 2} |u^\sigma
(x + \sigma (x,t),t)|^2
+ q^\sigma (x + \sigma (x,t),t) [\det((X^\sigma)^\prime_a
\circ X^{-1}) - D]\right\}\;,$$
where $D$ as before is given by (1.2) and satisfies (1.3).  Noting
that the composition of maps $X^\sigma$ and $X$ gives $(X^\sigma \circ
X^{-1})(x,t) = x +
\sigma  (x,t)$ we conclude with
$$L(\omega ) = \int dx \left\{{D\over 2} |u^\sigma (x + \sigma (x,t),t)|^2
+ q^\sigma (x + \sigma (x,t),t) [\det (I + \sigma^\prime_x) -
D]\right\}\;. \eqno(2.4)$$
At this stage we make the crucial assumption that $\sigma$ is sufficiently
small
that the following Taylor expansions may be truncated at linear order:
$$\eqalignno{
u^\sigma (x + \sigma (x,t),t) &\sim u(x,t) + (\sigma (x,t)\cdot
\nabla) u(x,t)\cr
q^\sigma (x + \sigma (x,t),t) &\sim q(x,t) + (\sigma (x,t) \cdot
\nabla) q(x,t) &(2.5)\cr}$$
where
$$\eqalignno{
u(x,t) &= \la u^\sigma (x + \sigma (x,t),t)\ra\,,\cr
q(x,t) &= \la q^\sigma (x + \sigma (x,t),t)\ra\,, &(2.6)\cr}$$
and $\la\cdot\ra$ denotes averaging with respect to the
random event $\omega $.  Thus at this level of
approximation (2.4) becomes
$$\eqalignno{
L(\omega ) &= \int dx \Big\{D\left[{1\over 2} |u(x,t)^2| + u(x,t)
\cdot (\sigma
(x,t) \cdot \nabla u(x,t)) + {1\over 2} |(\sigma (x,t) \cdot \nabla)
u(x,t)|^2\right]\cr
&+ [q(x,t) + (\sigma (x,t) \cdot\nabla)q(x,t)] [\det (I + \sigma
^\prime_x) - D(x,t)]\Big\}\;. &(2.7)\cr}$$
Therefore the averaged Lagrangian $\la L\ra$ is found to be
$$\eqalignno{
\la L\ra = \int dx \Big\{ {D\over 2} [|u|^2 &+ 2u \cdot
(\la\sigma \ra\cdot\nabla)u + \la\sigma_i\sigma_j
\ra \partial_i u\cdot \partial_j u] +\cr
+ q[\la\det(I + \sigma^\prime_x)\ra - D] &- D(\la\sigma
\ra \cdot\nabla)q + (\la\sigma \det(I + \sigma^\prime_x)
\ra\cdot\nabla) q\Big\}\;, &(2.8)\cr}$$
where we use the notation $\partial_i = {\partial\over\partial x_i},
i = 1,2,3$.  Then the variational derivatives of $\la L\ra$ are given by
$$\eqalignno{
{1\over D} {\delta\la L\ra \over \delta u} &=
(1 - {1\over D}\nabla\cdot (D\la\sigma \ra))u - {1\over D} \partial_i
(D\la\sigma_i\sigma _j\ra \partial_j u)\cr
{\delta\la L\ra \over \delta D} &= (1 +
\la\sigma \ra \cdot\nabla)q + {1\over 2} [|u|^2 + 2u \cdot
(\la\sigma\ra \cdot\nabla)u +
\la\sigma_i\sigma_j\ra (\partial_j u) \cdot (\partial_i u)]
= -Q &(2.9)\cr
{\delta\la L\ra \over \delta q} &= \la\det (I + \sigma
^\prime_x)\ra - D + \nabla \cdot (\la\sigma \ra D -
\la\sigma  \det(I + \sigma ^\prime_x)\ra)\;.\cr}$$
By stationarity of $\la L\ra$ under variations in $q$, the last equation in the
set (2.9) becomes
$$D = \la \det(I + \sigma ^\prime_x)\ra +
\nabla\cdot (\la\sigma \ra D) - \nabla \cdot \la\sigma \det(1 +
\sigma ^\prime_x)\ra\;.$$
In order for the mean flow $u$ to be incompressible, one takes $D =
1$.  This imposes the condition
$$1 = \la\det(I + \sigma ^\prime_x)\ra + \nabla\cdot
\la\sigma \ra - \nabla\cdot \la\sigma \det (I+
\sigma^\prime_x)\ra  \eqno(2.10)$$
on the statistics of the fluctuations.  Under this condition, the
Euler-Poincar\'e equation (1.4) and the equation (1.3) (for $\la
L\ra$ instead of $L$) can be written as
$${\partial\over\partial t}v + (u\cdot\nabla)v + v_j \nabla u_j +
\nabla Q = 0\;,\quad{\rm with}\quad \nabla\cdot u = 0\,, \eqno(2.11)$$
where we define
$$v
\equiv \bigg[{1\over D} {\delta\la L\ra \over \delta u}\bigg]_{D=1}
=\ (1 - \nabla\cdot\la\sigma \ra)u - \partial_i (\la\sigma
_i\sigma _j\ra\partial_j u)\,. \eqno(2.12)$$
These equations are slight generalizations of the $n$-dimensional Camassa-Holm
equations.  The latter correspond to the case where the isotropy
conditions
$$\la\sigma \ra = 0\;,\; \la\sigma _i\sigma _j\ra =
\alpha ^2\delta _{ij}\;, \eqno(2.13)$$
hold.  If moreover the statistics of $\sigma$ are homogeneous, then $\alpha
^2$ is
constant.  Under this form the equations (2.11), (2.12) were originally
derived.$^{14,15}$ That derivation generalizes a one-dimensional integrable
dispersive shallow water model studied in Camassa and Holm$^{16}$ to
$n$-dimensions and provides the interpretation of $\alpha
$ as the typical mean amplitude of the fluctuations as in (2.13).\smallskip

\noindent
{\bf Remark.} The ideal Camassa-Holm equations, or Euler alpha-model, in
(2.11) is formally the equation for geodesic motion on the
diffeomorphism group with respect to the metric given by the mean kinetic
energy Lagrangian $\la L\ra$ in equation (2.8), which is right-invariant
under the action of the diffeomorphism group. See Holm et al.$^{15}$ for
detailed discussions, applications and references to Euler-Poincar\'e
equations of
this type for ideal fluids and plasmas. After the original derivation of
equation
(2.11) in Euclidean space,$^{14,15}$ Holm et al.$^{17}$ and Shkoller$^{18}$
generalized it to Riemannian manifolds, discussed its existence and
uniqueness on
a finite time interval, and amplified the relation found earlier$^{14}$ of this
equation to the theory of second grade fluids.  Additional properties of
the Euler
equations, such as smoothness of the geodesic spray (the Ebin-Marsden
theorem) are
also known for the Euler-$\alpha$ equations and the limit of zero viscosity for
the corresponding viscous Navier-Stokes-$\alpha$ equations is known to be a
regular
limit, even in the presence of boundaries for homogeneous (Dirichlet) boundary
conditions.$^{17,18}$ Some of the most interesting solutions of the Euler
alpha-model could actually leave the diffeomorphism group due to a loss of
regularity. (This is seen in the one-dimensional Camassa-Holm equation.$^{16}$)
Such solutions may be interpreted in the sense of generalized flows, as done by
Brenier$^{19}$ and Shnirelman.$^{20}$ A functional-analytic study of the Euler
alpha-model is made in Marsden et al.$^{21}$\smallskip

\noindent
{\bf Adding viscosity.} We note that
$v$ in (2.12) represents a momentum.  Therefore we propose that the
viscous variant of (2.11) should take the following form, in which the
viscosity
acts to diffuse this momentum,
$${\partial\over\partial t} v + (u\cdot\nabla)v + v_j \nabla u_j
= \nu \Delta v - \nabla Q\;,\quad \nabla\cdot u = 0\;. \eqno(2.14)$$
Again, $v$ is given by (2.12).  Throughout we will refer to
equation (2.14) with definition (2.12) as the viscous Camassa-Holm
equations (VCHE), or Navier-Stokes alpha-model (NS-$\alpha$). The standard
Navier-Stokes equations are recovered when $\alpha$ is set to zero.
The VCHE (2.14) in three dimensions possesses global
existence and uniqueness, as well as a global attractor whose bounds on fractal
dimension show cubic scaling with domain size, as expected in the Landau
theory of
three-dimensional turbulence. The proofs of these properties of the VCHE, or
NS-$\alpha$ model, are given in Foias et al.$^{16f}$

Since in (2.14), $\sigma $ appears at power up to $2$ and we assume
$|\sigma |$ to be small (at least in average), the constraint (2.10)
can be given a simpler form by using the approximation
$$\eqalign{
&\la\det (I + \sigma ^\prime_x)\ra - 1 \sim
\nabla\cdot \la\sigma\ra  + \la\partial_1\sigma _1 \cdot
\partial_2\sigma _2 - \partial_2\sigma _1 \cdot \partial_1
\sigma_2\ra  +\cr
&+ \la\partial_2\sigma _2 \cdot \partial_3\sigma _3 - \partial_3\sigma
_2 \cdot \partial_2\sigma _3\ra + \la\partial_3\sigma _3 \cdot
\partial_1\sigma _1 - \partial_1\sigma _3 \cdot \partial_3
\sigma_1\ra\;.\cr}$$
Then (2.10) becomes (by neglecting the terms of degree $\geq 3$ in
$\sigma $)
$$\eqalignno{
\nabla &\cdot \la(\nabla\cdot\sigma)\sigma \ra -
\nabla\cdot\la\sigma \ra \sim \la \partial_1\sigma _1
\cdot \partial_2\sigma _2 - \partial_2\sigma_1 \cdot
\partial_1\sigma _2\ra +\cr
&+ \la \partial_2\sigma _2 \cdot \partial_3\sigma _3 -
\partial_3\sigma _2 \cdot \partial_2\sigma_3\ra +
\la \partial_3\sigma _3 \cdot \partial_1\sigma _1 -
\partial_1\sigma _3 \cdot \partial_3\sigma_1\ra\;. &(2.15)\cr}$$
See Gjaja and Holm$^{22}$ for the corresponding derivation of equations in
the form (2.11) in Generalized Lagrangian mean (GLM) theory with $\la\sigma
\ra=0$ and no viscosity.  We note that GLM theory provides no
closure.

\vskip18pt
\noindent 3.  \underbar{Connection with Continuum Mechanics}.  A mechanical
interpretation of these equations may be obtained by rewriting the VCHE (2.14)
(in the case where $\la\sigma\ra = 0, \alpha ^2 \equiv$ constant) in the
equivalent `constitutive' form
$${du\over dt} = \div\hbox{\bfmathit T}\;,\;\hbox{\bfmathit T} =
-p\hbox{\bfmathit I} + 2\nu (1 - \alpha ^2\Delta)\hbox{\bfmathit D} +
2\alpha ^2 \dot{\hbox{\bfmathit D}}\;, \eqno(3.1)$$
with $\nabla\cdot u = 0, \hbox{\bfmathit D} = (1/2)
(\nabla u + \nabla u^T), {\bf \Omega} =
(1/2) (\nabla u - \nabla u^T)$, and co-rotational (Jaumann)
derivative given by $\dot{\hbox{\bfmathit D}} = d\hbox{\bfmathit D}/dt
+ \hbox{\bfmathit D}{\bf \Omega} -
{\bf \Omega} \hbox{\bfmathit D}$, with $d/dt =
\partial/\partial t + u\cdot\nabla$.  In this
form, one recognizes the constitutive relation for VCHE as a variant
of the rate-dependent incompressible homogeneous fluid of second
grade,$^{23,24}$ whose viscous dissipation, however, is modified by
the Helmholtz operator $(1 - \alpha ^2\Delta)$.  Thus, the VCHE, or
NS-$\alpha$ closure model is not only Galilean invariant; it also satisfies
the continuum mechanics principles of objectivity and material frame
indifference. There is a tradition at least since Rivlin$^{25}$ of using
these continuum mechanics principles in modeling turbulence (see also
Chorin$^{26}$).  For example, this sort of approach is taken in deriving
Reynolds stress algebraic equation models.$^{27}$ Rate-dependent closure
models of
mean turbulence have also been obtained by the two-scale DIA approach$^{28}$
and by the renormalization group methods.$^{29}$
\vskip18pt

\noindent 4.  \underbar{Closure Ansatz}.  Since VCHE describe mean
quantities, we propose to use (2.14) as a
turbulence closure model and test this ansatz by applying it to
turbulent channel and pipe flows. For this purpose we also assume that as
long as
the boundary effects can be neglected, the isotropy conditions (2.13) hold.
It is
also appropriate to recall that the Reynolds equations are the averaged
Navier-Stokes equations$^{8,28}$
$${\partial\over\partial t} \bar u + (\bar u\cdot \nabla)\bar u = \nu
\Delta \bar u - \nabla \bar p - \overline{(u - \bar
u)\cdot\nabla(u - \bar u)}\,,
\quad \nabla\cdot \bar u = 0\,, \eqno(4.1)$$
where the upper bar denotes the ensemble average, $\bar u$ is the
mean flow, $\bar p$ the mean pressure and
$-\overline{((u-\bar u)\cdot\nabla)(u-\bar u)}$ is the divergence of
the Reynolds stresses.  Our ansatz asserts that:

\item{a)}  $\bar u$ is approximatively the solution $u$ of the VCHE
with the same symmetry and boundary conditions as $\bar u$.

\item {b)}  The Reynolds stress divergences are given by appropriate
terms in the VCHE found by matching equations (2.14) and (4.1).

\vskip18pt

\noindent 5.  \underbar{The Reynolds equations for channel flows}.
For turbulent channel flow (see, e.g., Townsend$^{30}$), the mean velocity in
(4.1) is of the form $\bar{\hbox{\bfmathit u}} = (\bar U(z), 0,0)^{tr}$,
with $\bar p =
\bar P(x,y,z)$ and the
Reynolds equations (4.1) reduce to
$$\eqalignno{
-\nu \bar U\spp &+ \partial_z \la wu\ra = -\partial_x \bar
P\;,\cr
\partial_z \la wz\ra &= -\partial_y \bar P\;,\quad
\partial_z \la w^2\ra = -\partial_z \bar P\;, &(5.1)\cr}$$
where $(u,v,w)^{tr} = \hbox{\bfmathit u} - \bar{\hbox{\bfmathit
u}}\;$  is the fluctuation of the velocity in the infinite channel
$\{(x,y,z) \in
\bR, -d \leq z \leq d\}$.  The (1,3) component of the averaged stress
tensor $T = - \bar p I - \overline{u\otimes u} + \nu (\nabla \bar u +
(\nabla \bar u)^{tr})$ is given by $\langle T_{13}\rangle = \nu \bar
U\sp(z) - \langle wu\rangle$.  At the boundary, the velocity
components all vanish and one has the stress condition
$$\mp\tau_0 = \langle T_{13}\rangle \big|_{z=\pm d} = \nu \bar U\sp
(z)\big|_{z=\pm d}\;, \eqno(5.2)$$
upon using $\langle wu\rangle = 0$ at $z = \pm d$.  Hence, the
Reynolds equations imply $\langle wv\rangle (z)\equiv 0$ and $\bar P
= P_0 - \tau_0 x/d - \langle w^2\rangle (z)$, with integration
constant $P_0$.

\vskip18pt

\noindent 6.  \underbar{The VCHE for channel flows}.  Passing to
the VCHE in the channel, we denote the velocity
$u$ in (2.14) by $\hbox{\bfmathit U}$ and seek its steady state
solutions in the form $\hbox{\bfmathit U} = (U(z),0,0)^{tr}$ subject to the
boundary
condition $U(\pm d) = 0$ and the symmetry condition $U(z) = U(-z)$.
In this particular case, the steady VCHE reduces to,
$$-\nu ((1 -\beta ^\prime)U)^\prime + \nu (\alpha ^2
U^\prime)^{\prime\prime\prime} = -\partial_x \tilde\pi\;,
\quad 0 = -\partial_y \tilde\pi\;,
\quad 0 = -\partial_z \tilde\pi\,, \eqno(6.1)$$
where $\alpha^2 = \la\sigma^2_3\ra, \beta = \la\sigma_3\ra$
and $\tilde\pi = \pi + \int (U - \beta \sp U -
(\alpha ^2 U\sp)\sp) Udz$.

In accord with the statistical assumptions in the Reynolds
equation, we also take the statistics of $\sigma $ to be invariant
under horizontal translations.  As already mentioned above,
we will suppose that away from the
wall, i.e. for $|z| \leq d_0$ with $0 < d_0 < d$ we have
$$\alpha (z) \equiv \alpha _0\;,\quad \beta (z)\equiv 0\;, \eqno(6.2)$$
with constants $d_0$ and $\alpha _0$ to be determined later.  The following
heuristic argument may provide some help in understanding this
length-scale $\alpha_0$.  Clearly $\alpha $ and $\beta $ must
depend on $d,\tau_0,\nu ,z$, the only physical quantities present.
Dimensional analysis then implies (with two suitable functions $f$
and $g$) that
$${\alpha\over d} = f\left(R_0, {d - |z| \over \ell_*}\right)\;,\quad
{\beta\over d} = g \left(R_0, {d - |z| \over \ell_*}\right)\;,
\eqno(6.3)$$
where $d - |z|$ is the distance to the wall, while
$$R_0 = \tau_0^{1/2}d/\nu \;,\quad \ell_* = d/R_0\;, \eqno(6.4)$$
i.e., $R_0$ is the wall-stress Reynolds number and $\ell_*$ is the
wall-length unit.  By eliminating $R_0$ in
(6.3) we can write
$${\alpha\over d}  = h\left( {\beta \over d}\;,\; {d - |z| \over \ell_*}\right)
\eqno(6.5)$$
with some function $h$ of two variables.  Assuming that $h(0,\infty
)$ exists and noticing that
$$h\left(0,{d - |z| \over \ell_*}\right) = h\left(0, {d - |z| \over
d} R_0\right)\;,$$
we obtain (as long as $|z| \leq d_0)$ that, for $R_0$ large enough,
the ratio
$${\alpha \over d} \equiv {\alpha _0\over d} \sim h(0,\infty )$$
is independent of $R_0$.  This heuristic prediction will be confirmed
later in a more rigorous way.

Finally, let us note that due to the symmetry of the physical
setting, we can also assume that
$$\sigma _3 (x,y,-z,t;\omega ) \equiv -\sigma _3(x,y,z,t;\omega )$$
and therefore
$$\beta (-z,t) \equiv -\beta (z,t)\;,\quad \alpha (-z,t)\equiv \alpha
(z,t)\;. \eqno(6.6)$$

\vskip18pt

\noindent 7.  \underbar{Realizability conditions}.  Recall that
the statistics of $\sigma $ are subjected to the condition
(2.15).  In the present case this takes the form
$$\displaylines{
\partial_3 \la(\nabla\cdot\sigma )\sigma _3\ra -
\partial_3\beta \cr
= \la(\partial_1\partial_3\sigma _1 +
\partial_2\partial_3\sigma _2)\sigma _3\ra +
\la(\partial_1\sigma _1 + \partial_2\sigma _2)\cdot
\partial_3\sigma _3\ra + {1\over 2} \partial_3^2 \alpha ^2
 - \partial_3\beta\cr
\sim \la\partial_1 \sigma _2 \cdot \partial_2\sigma _2 -
\partial_2\sigma _1 \cdot \partial_1\sigma _2\ra +
\la(\partial _1\sigma _1 + \partial_2\sigma _2) \cdot
\partial_3\sigma _3\ra - (\la\partial_3\sigma _2 \cdot
\partial_2\sigma _3\ra + \la\partial_1\sigma
_3\cdot\partial_3\sigma _1\ra )\;,\cr}$$
whence
$${1\over 2} (\alpha^2)\spp - \beta\sp \sim \langle\partial_1\sigma _1
\cdot \partial_2\sigma _2 - \partial_2\sigma _1 \cdot
\partial_1\sigma _2\rangle\;. \eqno(7.1)$$
The meaning of $\sigma $ forces
$$-d-z \leq \sigma _3 (x,y,z,t;\omega) \leq d-z\;\;{\rm for}\;\; |z|
\leq d\;. \eqno(7.2)$$
In this case one can prove that the following conditions hold

$$-d-z \leq \beta (z) \leq d-z\;,\quad\alpha (z)^2 \leq d^2 - z^2 -
2z\beta (z)\;\;{\rm for}\;\; |z|\leq d\;. \eqno(7.3)$$
Indeed, if $P = P_{z,t}$ denotes the probability distribution of
$\sigma _3(z,t;\omega )$ and
$$\beta ^+ = \int_{\{\sigma _3\geq 0\}} \sigma _3 P(d\sigma _3)\;,\quad
\beta ^- = \int_{\{\sigma _3<0\}} |\sigma _3| P(d\sigma _3)\,,$$
then
$$\beta  = \langle\sigma _3\rangle = \beta ^+ - \beta ^-\;,\quad  \beta ^- \leq
(d+z)P(\{\sigma _3 < 0\})\;,\quad \beta ^+ \leq (d-z)P(\{\sigma _3 \geq
0\})\,.$$
Thus,
$$(d+z)^{-1} \beta ^- + (d-z)^{-1} \beta ^+ \leq 1\;,$$
so that
$$2d\beta _- \leq d^2 - z^2 - (d+z)\beta \;.$$
On the other hand,
$$\eqalign{
\alpha ^2 = \langle\sigma ^2_3\rangle = \int \sigma ^2_3 P(d\sigma
_3) &\leq (d + z)\beta ^- + (d-z)\beta ^+ \leq\cr
&\leq 2d\beta ^- + (d-z)\beta  \leq d^2 - z^2 - 2z\beta \;.\cr}$$
This establishes the second inequality in (7.3).  The first one is
obvious.

The Cauchy-Schwarz inequality produces the supplementary constraint
$$|\beta (z)| \leq \alpha (z)\;{\rm for}\; |z|\leq d\;. \eqno(7.4)$$
It is easy to check that the conditions (7.2) and (7.4) are also
sufficient for the existence of a random variable $\sigma
_3(x,y,z;\omega )$ satisfying (6.6) and (7.3) and statistically
depending only on $z$.  For any such $\sigma _3$, choose some
homogeneous random vector $[\sigma _1^0(x,y),\sigma _2(x,y)]$ such
that $\gamma= \langle\partial_1\sigma _1^0 \cdot \partial_2\sigma
_2 - \partial_2\sigma _1^0 \cdot \partial_1\sigma _2\rangle \ne 0$.
Set $\sigma _1 = {(2\gamma)^{-1} }\big((\alpha ^2)\spp - 2\beta
\sp\big)\sigma _1^0$.  Then $\sigma  = (\sigma _1,\sigma _2,\sigma _3)$
has all the required statistical properties.  We conclude that the
inequalities (7.3) and (7.4) are the realizability conditions for the
lengths $\alpha $ and $\beta $ in the VCHE (6.1).

\vskip18pt

\noindent 8.  \underbar{Comparing VCHE with the Reynolds equation}.
Comparing (5.1) and (6.1), we identify counterparts as,
$$\eqalignno{
\bar U = U\;,\; \partial_z\la w u\ra &= \nu [(\alpha
^2U\sp)^{\prime\prime\prime} - (\beta\sp U)\spp] + p_0\;,\cr
\partial_z \la w v\ra = 0\;,\; \nabla &(\bar P +
\la w ^2\ra) = \nabla(\tilde\pi - p_0x)\;, &(8.1)\cr}$$
for a constant $p_0$.  This identification gives
$$\eqalignno{
&\la w v\ra (z) = 0\,,\cr
- \la w u\ra (z) = -p_0z &- \nu [(\alpha ^2 U\sp)\spp(z)
- (\beta \sp U)\sp(z)]\,, &(8.2)\cr}$$
and leaves $\la w^2\ra$ undetermined up to an arbitrary
function of $z$.  A closure relation for $-\langle w u\rangle$
involving the third derivative $U^{\prime\prime\prime}(z)$ also appears in
Yoshizawa,$^{28}$ cf. equation (8) of Wei and Willmarth.$^{4}$

>From (6.1) it follows that $\partial _x\tilde\pi = \pi_2$ is constant.
Therefore integrating twice in $z$, the first equation in (6.1) gives
$$-\nu (1-\beta \sp(z))U(z) + \nu (\alpha ^2(z)U\sp(z))\sp = \pi_0
+ \pi_1 z - {1\over 2} \pi_2 z^2 \eqno(8.3)$$
with constants $\pi_i(i= 0,1,2)$.  But the left hand side of (8.3) is
symmetric under the change $z\mapsto -z$, so $\pi_1 = 0$ and we obtain the
following relation among the profiles of $\beta(z)$, $\alpha(z)$ and $U(z)$,
$$-\nu (1-\beta \sp(z))U(z) + \nu (\alpha (z)^2 U\sp(z))\sp = \pi_0 -
{1\over 2} \pi_2 z^2\;\;{\rm for}\;\; |z| \leq d\;. \eqno(8.4)$$
For $|z| \leq d_0, \beta(z) \equiv 0, \alpha (z) \equiv \alpha
_0 > 0$ and (8.4) becomes
$$-U(z) + \alpha _0^2 U\spp(z) = {1\over \nu }\pi_0 - {1\over 2\nu }
\pi_2 z^2\;\;{\rm for}\;\;|z| \leq d_0\;. \eqno(8.5)$$
Since $U$ is symmetric in $z$, we obtain
$$U(z) = a \left(1 - {\cosh (z/\alpha _0) \over \cosh (d_0/\alpha
_0)}\right) + b\left(1 - {z^2\over d_0^2}\right) + c\;\;{\rm for}\;\;
|z| \leq d_0\;, \eqno(8.6)$$
where the constants $a,b$ and $c$ satisfy the conditions
$$c = U(\pm d_0)\;,\;\pi_0\nu  = -a - b(1 + 2\alpha _0^2/d_0^2) - c\;,
\;\pi_2\nu = -2b/d_0^2\;. \eqno(8.7)$$
It is worth mentioning here that with an antisymmetry condition for
$U(z)$ and with (8.6) changed accordingly, one may address turbulent
shear flows (Couette flows) by the same analysis as developed in this
paper.

Integrating (8.4) on $[-d,0]$ gives
$$-\nu  \int^0_{-d} (U(z) + \beta (z)U\sp(z))dz - \alpha (-d)^2\tau_0 =
\pi_0d - {1\over 6} \pi_2d^3 \eqno(8.8)$$
where we used (5.2) as well as $U\sp(0) = 0, \beta(0) = 0$ and
$U(-d)=0$.  Denoting
$$\eqalignno{
\Uav &= {1\over 2d} \int^d_{-d} U(z)dz = {1\over d} \int^0_{-d}
U(z)dz =\cr
&= {1\over d} \int^{-d_0}_{-d} U(z)dz + \left[a\left(1 - {\alpha _0\over
d_0}\tanh {d_0\over \alpha _0}\right) + {2\over 3}b + c\right]
{d_0\over d} &(8.9)\cr}$$
allows (8.8) to be written also as
$$-\nu d\Uav - \nu  \int^{d_0}_{-d} \beta (z) U\sp (z)dz - \alpha
(-d)^2 \tau_0 = \pi_0d - {1\over 6} \pi_2 d^3\;. \eqno(8.10)$$

\vskip18pt

\noindent 9.  \underbar{Empirical qualitative properties}.  It is
universally accepted that the maximum of $U$ is at $z = 0$
(i.e. the center of the channel) and that $U\sp(z)\cdot z < 0$ for
$0<|z|<d$.  Also all experimental data show that $U\spp(z) < 0$ over
most of the channel.  Thus
$$R := {d\over \nu } \Uav = {1\over 2\nu } \int^d_{-d} U(z)dz \leq
R_c \equiv {d\over \nu } U(0) \eqno(9.1)$$
and (using the concavity property of $U$)
$$R \geq {1\over \nu}  \int^0_{-d} {z+d \over d} U(0) dz = {1\over 2}
R_c\;. \eqno(9.2)$$
%
Then (9.1), (9.2) can be given the form
$${1\over 2} {U(0) \over u_*} \leq {R\over R_0} \leq {U(0)\over u_*}\;.
\eqno(9.3) $$
All the empirical evidence shows that
$${R\over R_0^2} \ll 1 \ll {R\over R_0}\;\;{\rm for}\;\; R_0\gg 1\;.
\eqno(9.4)$$
Throughout, the properties (9.3) and (9.4) will be taken as granted.

\vskip 18pt

\noindent 10.  \underbar{The wall units representation}.  In the
lower half of the channel, the mean velocity $U$ can be
expressed in wall units using the notation $\phi(\eta) =
U(z)/u_*,\eta = (z+d)/\ell_*$, with $\ell_* = \nu /u_* = d/R_0$.  In
this representation, (8.6) becomes
$$\eqalignno{
\phi(\eta) &= {a\over u_*} \left(1 - {\cosh\xi(1-\eta/R_0) \over
\cosh\xi(1-\eta_0/R_0)}\right) \cr
&+ {b\over u_*} \left(1 - \left({1-\eta/R_0 \over
1-\eta_0/R_0}\right)^2\right) + \phi(\eta_0)\;, &(10.1)\cr}$$
for $\eta_0 \leq \eta \leq R_0$, where $\xi = d/\alpha$ and $\eta_0 =
(d - d_0)/\ell_* \sim \alpha _0/\ell_* = R_0/\xi$.

The definition of $\phi$, implies $R = \int^{R_0}_0
\phi(\eta)d\eta$.  Hence (10.1) gives
$$R = {a(R_0-\eta_0)\over u_*} \left(1-{\tanh\xi(1-q_0) \over
\xi(1-q_0)}\right) + {2b(R_0-\eta_0) \over 3u_*} + \phi(\eta_0)
(R_0 - \eta_0) +
\int^{\eta_0}_0 \phi(\eta)d\eta\;.$$
To conclude this computation it is sufficient to approximate $\phi$
on $(0,\eta_0)$ by the piecewise linear function equal to $\eta$ for
$0 < \eta\leq \eta_*$ and $\phi_0 + (\eta - \eta_0)\phi\sp_0$ for
$\eta_* \leq \eta \leq \eta_0$, where $\phi_0 = \phi(\eta_0),
\phi\sp_0 = \phi\sp(\eta_0)$ and $\eta_* = (\phi_0 - \eta_0
\phi\sp_0)/(1-\phi\sp_0)$.  We obtain
$$\eqalignno{
{R\over R_0} \approx {a(1-q_0)\over u_*}
&\left(1 - {\tanh\xi(1-q_0)\over
\xi(1-q_0)}\right) + {2b(1-q_0)\over 3u_*}\cr
+ (1-q_0)\phi_0 &+ (1-\phi\sp_0)^{-1} \left(\phi_0q_0 -
{q_0^2R_0\phi\sp_0
+ \phi^2_0/R_0\over 2}\right) &(10.2)\cr}$$
where
$$\phi\sp_0 = (a/u_*) (\xi/R_0)\tanh(\xi(1-q_0)) + 2(b/u_*)/R_0(1-q_0)\;.
\eqno(10.3)$$
Using this and solving for $\phi_0$ gives an explicit function
$\phi_0 = \phi_0(q_0;R,R_0;a/u_*, \hfill\break
b/u_*;\xi)$, namely
$$\phi_0(q_0;R,R_0;a/u_*,b/u_*;\xi) = R_0\Bigg\{B-\sqrt{B^2-{1\over
R_0}[2(R/R_0-C) + q_0^2R_0\phi^\prime_0]}\Bigg\} \eqno(10.4)$$
where $\phi\sp_0$ is given by (10.3),
$$\eqalignno{
B &= (1-q_0) (1-\phi\sp_0) + q_0\cr
C &= {a(1-q_0)\over u_*}
\left(1- {\tanh\xi(1-q_0) \over \xi(1-q_0)}\right)
+ {2\over 3} {b(1-q_0)\over u_*} &(10.5)\cr}$$
and the choice of the root $\phi_0$ in (10.2), (10.3) will be
justified at the end of Section 11.

Thus (10.1) becomes
$$\eqalignno{
\phi(\eta) &= {a\over u_*} \left(1- {\cosh\xi(1-\eta/R_0) \over
\cosh\xi(1-q_0)}\right) +&(10.6)\cr
&+ {b\over u_*} \left(1-\left({1-\eta/R_0\over 1-q_0}\right)^2\right)
+ \phi_0(q_0; R,R_0; {a\over u_*}, {b\over u_*};\xi)\;\;{\rm
for}\;\; q_0R_0\leq\eta\leq R_0\;.\cr}$$
In (10.6) the constants $a/u_*, b/u_*,\xi$ and $q_0$ may depend on
$R_0$.  As we will show below, Nature seems to choose them as constants
(at least for large $R_0$).  Recall that in Section 6 we already gave
a heuristic argument that $\xi = d/\alpha $ should be independent of
$R_0$ if $R_0$ (or $R$) is large enough.

\vskip18pt

\noindent 11.  \underbar{The off wall region}.  The empirical data
up to now suggest that for a fixed channel there
is a range $(z_1,z_2)$ (with $z_1z_2> 0)$ inside the channel such
that for $z$ in that range, the von K\'arm\'an log-law is a good
approximation to $U(z)$, at least for $R$ (or $R_0$) large enough.
Since for those $z$ we have
$$U(z_2) - U(z) = {1\over \kappa }\ln {z_2\over z} = {1\over \kappa }
\left(\ln{z_2\over d} - \ln{z\over d}\right)$$
(where $\kappa\sim .4$ is the von K\'arm\'an constant), $U(z_2) - U(z)$ is
a function of $z/d$ only (i.e. independent of $R_0$).  We will posit
now the following weaker condition.

For $R$ (or $R_0$) large enough, there exists a fixed range $(z_1/d,
z_2/d)$ such that for $z/d$ in that range, $U(z_2) - U(z)$ is a
function of $z/d$, independent of $R_0$.

Note that we make no assumption on the length of the range.  The
classical ``defect law'' of Izakson, Millikan and von Mises$^{31}$
(pp.~186--188) is the
particular case of our condition when one of $z_i$'s is $0$, and the
range is assumed to be wide.

Passing to the wall units representation we can formulate our
assumption as:  There exists $0<q_1<q_2<1$, such that for $q_1R_0
\leq \eta\leq q_2R_0, \phi(\eta_2) - \phi(\eta)$ is a function of $q
= \eta/R_0$ only.  Since we expect $q_0$ in (10.6) to be quite small,
we will take $q_0\leq q_1$.

We will prove now that under the above conditions, there exist
absolute constants $a_*, b_*$ and $\xi_*$ such that
$$a \sim a_*u_* \cosh \xi_*(1-q_0)\;,\; b \sim b_*u_*(1-q_0)\;\;{\rm
and}\;\; \xi=d/(\alpha _0 \xi_*) \eqno(11.1)$$
where $a,b,\xi$ and $q_o$ are as in (10.6).

Indeed let $f$ be the function defined by
$$f(q) = \phi(q_2R_0) - \phi(qR_0)\;\;{\rm for}\;\; q_1\leq q\leq q_2\;.
\eqno(11.2)$$
Then since $q_0 \leq q_1$ we have from (10.6)
$$f(q) = a_0[\cosh\xi(1-q) - \cosh\xi(1-q_2)] + b_0[(1-q)^2 -
(1-q_2)^2] \eqno(11.3)$$
where
$$a_0 = (a/u_*)/\cosh\xi(1-q_0)\;,\; b_0 = (b/u_*)/(1-q_0)^2\;.
\eqno(11.4)$$
Writing (11.3) for $q=q_1$, we obtain
$$b_0 = {f(q_1) - a_0[\cosh\xi(1-q_1) - \cosh\xi(1-q_2)] \over
(1-q_1)^2 - (1 - q_2)^2}\;.$$
Then (11.3) becomes
$$a_0g(\xi,q) = h(q)\;\;{\rm for}\;\; q_1\leq q\leq q_2 \eqno(11.5)$$
where, with $c_0$ an absolute constant,
$$\eqalignno{
g(\xi,q) &= \cosh\xi(1-q) - \cosh\xi(1-q_2) - c_1[(1-q)^2 -
(1-q_2)^2]\cr
h(\xi) &= f(q) - c_0[(1-q)^2 - (1-q_2)^2]\;, &(11.6)\cr}$$
and $a_0,c_1\xi$ are parameters, constant in $q$ but which may
depend continuously on $R_0$.  Note that
$$g(\xi,q_i) = h(q_i) = 0\;(i=1,2)\;,\; g(\xi,q)< 0\;\;{\rm for}\;\;
q_1 < q< q_2\;,\;\xi> 0\;. \eqno(11.7)$$
Thus (with $\bar q = (q_1 + q_2)/2)$
$$a_0 = h(\bar q)/g\!(\bar q)\;, \eqno(11.8)$$
and
$$g(\xi,q) h\!(\bar q) =
h(q)g\!(\xi,\bar q)\;\;{\rm for}\;\;q_1\leq q\leq
q_2\;. \eqno(11.9)$$
If $\xi = \xi(R_0)$ were not constant, then (11.9) would hold for $\xi$
in an interval $[\xi_1,\xi_2]$ with $0<\xi_1 < \xi_2$.  Differentiating (11.9)
with respect to $\xi$ gives
$$g^\prime_\xi (\xi,q) g\!(\xi,\bar q) =
g^\prime_\xi\!(\xi, \bar q) g(\xi,q)\,,$$
for $\xi_1\leq \xi\leq \xi_2, q_1\leq q\leq q_2$.  Introducing $\zeta
= \xi(1-q)$, it follows that
$$\sinh \zeta = h_0(\xi) + h_1(\xi)\zeta^2 + h_2(\xi) \cosh\zeta
\;\;{\rm for}\;\; \xi_1\leq\xi<\xi_2,\quad \xi(1-q_2) < \zeta <
\xi(1-q_1)\,,$$
where $h_0,h_1,h_2$ are explicit functions of $\xi$ only.  Clearly
this is impossible.

We conclude from this contradiction that there are absolute constants $q_0,
a_*,
b_*$ and
$\xi_*$ such that
$$\eqalignno{
\phi(\eta) &= a_*\left[\cosh\xi_*(1-q_0) - \cosh\xi_*\left(1- {\eta\over
R_0}\right)\right] \cr
&\qquad+ b_* \left[(1-q_0)^2 - \left(1-{\eta\over R_0}\right)^2\right] \cr
&\qquad+ \phi_0 (q_0; R,R_0;a_*\cosh\xi_*(1-q_0), b_*(1-q_0)^2;\xi_*)\,,
&(11.10)\cr}$$
for $q_0R_0 \leq \eta\leq R_0$,
where the function $\phi_0$ (see (10.4)) actually depends only
on $q_0, R_0$ and $R$.

The formula (11.10) can be also written as
$$\phi(qR_0) = \phi_1(q_0;q) + \phi_0 (q_0;R,R_0)\;\;{\rm for}\;\;
q_0\leq q\leq 1\,, \eqno(11.11)$$
where
$$\eqalignno{
\phi_1(q_0;q) &= a_* \cosh \xi_* (1-q_0) \left(1 - {\cosh\xi(1-q)\over
\cosh\xi(1-q_0)}\right) \cr
&\qquad+\ b_*(1-q_0)^2 \left(1 - \left({1-q\over
1-q_0}\right)^2\right) \;\;{\rm for}\;\; q_0\leq q\leq 1\;,\cr
\phi_0 (q_0;R,R_0) &=
\phi_0(q_0;R,R_0;a_*\cosh\xi_*(1-q_0),b_*(1-q_0)^2; \xi_*)\;.
&(11.12)\cr}$$
For $R_0\to\infty $ from (10.4) and (10.5) we have
$$\phi(q_0R_0) = \phi_0(q_0;R,R_0) \sim {R\over R_0} - C + {1\over
2} q_0^2 C_0\;, \eqno(11.13)$$
where $C$ is defined in (10.5) and $C_0 = R_0\phi^\prime_0$ is
constant according to (10.3).

We can now explain the choice of $\phi_0$ in (10.4).  The other
possible choice was
$$R_0\left\{ B + \sqrt{B^2 - {1\over R_0} [2(r/r_0 - C) + q^2_0 c_0]}
\;\right\}$$
which would have given for $R_0$ large enough
$$\phi(q_0R_0) \sim 2R_0$$
and consequently $\phi(R_0) \geq R_0$ which is contrary to the
established facts (9.3), (9.4).

\vskip 18pt

\noindent 12.  \underbar{The mean velocity profile
in the channel}.
Comparing the profile given by formula (11.10)
with an experimental mean velocity profile,
enables us to obtain the values $a_*, b_*$ and $\xi_*$ as well as the
smallest acceptable value $q_*$ for $q_0$.
In Figure 1,
we compare our formula with experimental data$^{4}$
for the Reynolds numbers $R_0$ equal to 714, 989, and 1608.
As these Reynolds numbers are small,
$a_*$ and $b_*$ have not reached their asymptotic values.
It appears, however, that $\xi_*$ has reached its asymptotic value.
We therefore allow $a_*$ and $b_*$ to vary slightly with $R_0$
while holding $\xi_*$ constant to fit the data.
It turns out that $\xi_*=35$ and $q_*=1/\xi_*$.
Note that this choice of $q_*$ corresponds exactly to
the condition that $|d-d_0|=\alpha$.


\vskip18pt

\noindent 13.  \underbar{The Reynolds shear stress}.  The shear
Reynolds stress is $-\langle uw\rangle$ (see Section 5).
Since $\la uw\ra\big|_{z=\pm d} = 0$, one must have
$$-\la uw\ra (z) = -\tau_0 {z\over d} - \nu \bar U\sp
(z)\;\;{\rm for}\;\; |z|\leq d\;. \eqno(13.1)$$
On the other hand $\bar U\equiv U$ and $-\la uw\ra$ is also
given by (8.2) with an appropriate constant $p_0$.  For $|z|\leq
d_0$, since $\alpha (z)\equiv \alpha _0 = 1/\xi_*,\beta (z)\equiv 0$, (8.2)
reduces to
$$-\la uw\ra (z) = p_0z - \nu \alpha ^2_0 U^{\prime\prime\prime}(z)
\eqno(13.2)$$
and
$$\nu \alpha ^2_0 U^{\prime\prime\prime}(z) = \nu  U\sp(z) - \pi_2
z\;.$$
Introducing this in (13.2) we see that (13.1) and (13.2) are
compatible if $p_0$ is given by
$$p_0 = -\pi_0 - \tau_0/d\;. \eqno(13.3)$$
Taking the wall units representation in (13.1) we obtain our
theoretical Reynolds shear stress
$$\eqalignno{
&{-\la uw\ra\over \tau_0} = 1 - {\eta\over R_0} - \phi\sp(\eta) =\cr
&= 1 - {\eta\over R_0} + {a_*\over R_0} \xi_* \sinh \xi_*
\left(1-{\eta\over R_0}\right) + {2b_*\over R_0}
\left(1-{\eta\over R_0}\right)\;\;{\rm for}\;\; q_* R_0\leq\eta\leq
R_0\;. &(13.4)\cr}$$

Figure~2 compares the corresponding experimental and theoretical
Reynolds shear stresses.
We use the same values for $a_*$, $b_*$ and $\xi_*$ as before.
The agreement in shear stresses does not extend as close to the wall
as the mean velocity profiles did.
The empirical matching of the mean velocity profiles as well as the
Reynolds shear
stresses are both given with $q_*=\sqrt 3/\xi_*$. We note that the
consistency of
this closure and the experiments found in the trends followed by the
Reynolds-stress profiles in Figure~2 is an exacting test of the fidelity of the
mean velocity profiles as well as a test of the Reynolds stress relation
predicted
by equation (13.4).

\vskip 18pt

\noindent 14.  \underbar{The near wall region}.  As already
mentioned above, in the near wall regions (i.e. where $0\leq
\eta\leq\eta_0 = q_*R_0$ and $2R_0 - q_*R_0\leq \eta\leq 2R_0),
\;\beta $ may be non-zero and $\alpha $ may depend (as does
$\beta $) on $\eta$ and $R_0$.  The VCHE (8.4) in the wall
units representation takes the form
$$\eqalignno{
(1-R_0\tilde\beta^\prime(\eta))\phi(\eta) &- R_0^2(\tilde\alpha
(\eta)^2\phi\sp(\eta))\sp = f_0 + 3f_1 \left(1- {\eta\over
R_0}\right)^2 \cr
&{\rm for}\;\; 0\leq\eta\leq R_0 &(14.1)\cr}$$
i.e., in the whole lower half of the channel.  In (14.1) we used the
notations
$$\tilde\alpha (\eta) = \alpha (z)/d\;,\; \tilde\beta (\eta) = \beta
(z)/d \eqno(14.2)$$
where $d + z = \eta\ell_*, \ell_* = d/R_0$, and
$$\eqalignno{
&f_0 = -\pi_0/\nu u_* = a_*\cosh\xi_* (1-q_*) + b_* \left[(1-q_*)^2
+{2\over\xi^2_*}\right] + \phi(q_*R_0) \cr
&f_1 = \pi_2d^2/\nu u_* = -b_*/3 &(14.3)\cr}$$
Of course we have
$$\tilde\alpha (\eta) = 1/\xi_*\;,\; \tilde\beta (\eta) = 0\;\;{\rm
for}\;\; q_*R_0\leq\eta\leq R_0\;, \eqno(14.4)$$
but the VCHE (14.1) does not define $\tilde\alpha (\eta),\tilde\beta
(\eta)$ near the wall (i.e. for $0\leq\eta\leq q_*R_0$).  However,
(14.1) gives some qualitative information on the behavior of
$\tilde\alpha $ and $\tilde\beta $ in the near wall region.  Indeed
integrating (14.1) we obtain
$$\eqalignno{
&{1\over R_0} \int^{R_0}_\eta \phi(\eta\sp)d\eta\sp +
\int^{R_0}_\eta \tilde\beta (\eta\sp)\phi\sp(\eta\sp)d\eta\sp +
\tilde\beta (\eta) \phi(\eta) + R_0\tilde\alpha (\eta)^2
\phi\sp(\eta) =\cr
&= f_0\left(1 - {\eta\over R_0}\right) + f_1\left(1 -
{\eta\over R_0}\right)^3 &(14.5)\cr}$$
for all $0\leq \eta\leq R_0$.  Thus (using $|\tilde\beta | \leq 2$,
see(7.3)) we find
$$\tilde\alpha (\eta)^2\phi'(\eta) \leq {1\over R_0} (f_0 + f_1 +
4\phi(\eta)) = O\left({R\over R_0^2}\right)$$
which in turn goes to zero when $R_0\to\infty $, by virtue of (9.4).
Thus, for $\eta$ fixed,
$$\tilde\alpha (\eta)^2\phi\sp(\eta)\to 0\;\;{\rm for}\;\;
R_0\to\infty\;. \eqno(14.6)$$
In particular, (using $\phi\sp(0) = 1$ and (9.3)) we obtain
$$\tilde\alpha (0)\to 0\;,\; \tilde\beta (0)\to 0\;\;{\rm for}\;\;
R_0\to\infty  \eqno(14.7)$$
and, due to (14.4),
$$\phi\sp(q_* R_0)\to 0\;\;{\rm for}\;\; R_0\to\infty\;. \eqno(14.8)$$
Moreover, writing (14.5) for $\eta = 0$,
$${R\over R_0} + \int^{R_0}_0 \tilde\beta (\eta\sp) \phi\sp (\eta\sp)
d\eta\sp + R_0 \tilde\alpha (0)^2 = f_0 + f_1 \eqno(14.9)$$
and subtracting (14.5) from (14.9) we obtain
$$\eqalignno{
{1\over R_0} \int^\eta_0 \phi(\eta\sp)d\eta\sp &+ \int^\eta_0
\tilde\beta (\eta\sp)\phi\sp(\eta\sp)d\eta\sp - \tilde\beta
(\eta)\phi(\eta) + R_0[\tilde\alpha (0)^2 - \tilde\alpha
(\eta)^2\phi\sp(\eta)] =\cr
&= f_0{\eta\over R_0}
+ f_1 \left[1-\left(1- {\eta\over R_0}\right)^3\right]
&(14.10)\cr}$$
Fixing $\eta$ and letting $R_0\to\infty $, from (14.10) we infer that
$$\int^\eta_0 \tilde\beta (\eta\sp)\phi\sp(\eta\sp)d\eta\sp -
\tilde\beta (\eta)\phi(\eta) + R_0[\tilde\alpha (0)^2 - \tilde\alpha
(\eta)^2\phi\sp(\eta)]\to 0\;. \eqno(14.11)$$
But for $\eta$ fixed such that $\phi\sp(\eta)$ remains away from $0$
when $R_0\to\infty$, we have $\beta (\eta)\to 0$ because of (7.4).
By virtue of Lebesgue's dominated convergence theorem we conclude
that
$$R_0[\tilde\alpha (0)^2 - \tilde\alpha (\eta)^2 \phi\sp(\eta)]\to
0\;\;{\rm when}\;\; R_0\to\infty \, \eqno(14.12)$$
provided that $\phi\sp(\eta)$ stays away from $0$.

It is instructive to connect (14.12) with the functions $g$ and $h$
considered in Section 6.  First we recall that
$$g(R_0,\eta) = \tilde\beta (\eta)\;,\quad f(\tilde\beta (\eta),\eta) =
\tilde\alpha (\eta)\,. \eqno(14.13)$$
Assuming that $\tilde\beta $ and $\tilde\beta \sp$ are continuous
across $\eta_0 = q_*R_0$ allows us to conclude that
$$g(R_0,R_0q) = -\int^{q_*}_q (q\sp - q_*)
\Big({\partial\over\partial q\sp}\Big)^2\!
g(R_0,R_0q\sp)dq\sp\;\;{\rm for}\;\; 0\leq q\leq q_*\;.$$
Since $q_*$ is small, assuming that
$$\bigg[\Big({\partial\over\partial
q'}\Big)^2\!  g(R_0,R_0q\sp) \bigg]
\bigg|_{q\sp=q_*-0}
= \gamma(R_0)$$ exists and
is not zero, we obtain that
$$g(R_0,\eta) \sim {1\over 2} \gamma(R_0)(q_* - \eta/R_0)^2\;.
\eqno(14.14)$$
But $\tilde\beta (0)\geq 0$, so $\gamma(R_0) > 0$.  If $\gamma(R_0) =
0$, we can proceed in a similar way by involving a higher derivative
of $g$ at $\eta_0$ from the left.  In all cases we have ended with a
representation
$$g(R_0,\eta)\sim g_0(R_0)g_1(\eta/R_0)\;\;{\rm for}\;\;
0\leq\eta\leq q_*R_0 \eqno(14.15)$$
where $g_0 \geq 0$, $g_1\geq 0$ and $g_1(q)$ is a decreasing function
of $q$ with $g_1(q_*) = g_1^\prime(q_*) = 0$.  From (14.15), (14.13)
and (14.12) we now obtain
$$h^2(g_0(R_0)g_1(0),0) - h^2(g_0(R_0)g_1(\eta/R_0),\eta)
\phi\sp(\eta) \sim 0 \eqno(14.16)$$
for $R_0$ large enough.  In (14.16), $g_1(\eta/R_0) \to g_1(0)$ for
$R_0\to\infty $ and $\eta$ fixed.  These arguments suggest that
$$h(\beta ,\eta) \sim h(\beta ,0)/\sqrt{\phi\sp(\eta)} \eqno(14.17)$$
provided $\beta $ is small and $1/\phi\sp(\eta)$ is bounded when
$R_0\to\infty $.  It is not clear if assuming equality in (14.17) is a
judicious
approximation of the function $h$.

A major difficulty in fine tuning our approach near the wall
resides in the unavailability of experimental data in the near wall region for
large Reynolds numbers.  To test whether the VCHE (14.5) is still valid
in this region we extrapolated the experimental profiles in Figure 1
into the near wall region according to Panton$^{9}$ to
obtain $\alpha$ from (13.5).   For simplicity of the graph, we will display
the $\alpha$ profile only for $R_0=1608$, which is
the highest Reynolds number in Figure~1.
As illustrated in Figure~3,
we find that the realizability conditions (in Section
7) are satisfied for appropriate choice of $\gamma(R_0)$ in (14.14) and
(14.13).
Clearly $\alpha$ lies between the upper constraint (7.3) and
the lower constraint (7.4) in the near wall region.
Thus, our basic ansatz is consistent with Panton's theory.

\vskip18pt

\noindent 15.  \underbar{Pipe flows and prediction}.  All the preceding
considerations on turbulent channel flows can be
suitably applied to turbulent pipe flows.  The substantial difference
between the mathematical treatment of the two types of flows, is that
for pipes, the cosh function is replaced by the first modified Bessel
function$^{32}$
$$I_0(r) = \sum^\infty _{n=1} {1\over (n!)^2}
\left({r^2\over 4}\right)^n\;. \eqno(15.1)$$
For instance the basic formula (10.11) becomes
$$\eqalignno{
\phi(\eta) &= {a\over u_*} \left[1 - {I_0(\xi(1-\eta/R_0) \over
I_0(\xi(1-\eta_0/R_0)}\right] + {b\over u_*} \left(1 -
\left({1-\eta/R_0\over 1 - \eta_0/R_0}\right)^2\right)\cr
&+ \phi(\eta_0)\;\;{\rm for}\;\; \eta_0 \leq \eta \leq R_0\;.
&(15.2)\cr}$$

For pipe flows, experimental data for quite large Reynolds numbers
are available (see Zagarola$^{7}$).
For these Reynolds numbers it is reasonable to assume
that $a_*$, $b_*$, and $\xi_*$ have each reached their
asymptotic values.
In Figure 4 we compare our profiles with experimental data of Zagarola.$^{7}$
We obtain the $a_*,b_*,\xi_*$ and $q_*$ by using the experimental data
for $R = 98,812$ and use the von K\'arm\'an drag law,
$R/R_0 \sim \log R_0$, to obtain
profiles for $R = 3,089,100$ and $35,259,000$. See also Chen et al.$^{33}$
for additional discussion and numerical details for these comparisons. 

We note that our
predictions are consistent with the von K\'arm\'an log law,$^{34}$ the
Barenblatt-Chorin power law,$^{10}$ as well as
with the presence of the `chevron' near the
center of the flow.  Our approach shows a logarithmic profile for
0.02 $R_0\leq \eta\leq 0.2 R_0$ and a chevron near the center of the
channel.  The Barenblatt-Chorin power law$^{10}$ may represent the
transition in the profile from the log law to the chevron.  Although
our approach is in good agreement with the experiments,$^{7}$ we
note that it has been argued that the experimental
mean velocity profiles are too low for high Reynolds numbers.$^{10}$
Finally, we observe that the chevron may reflect the
fact that, on the attractor of the
dynamical system in the phase space of the turbulent flow,
the Poiseuille-Hagen flow is recurrent.

\vfill\eject

\noindent\underbar{References}
\bibliography

\ref{[1]}  H. Reichardt,
``Messungen turbulenter Schwankungen,''
Die Naturwissenschaften {\bf 24/25} (1938),
404--408.\par

 \ref{[2]}  H. Eckelmann,
``Experimentelle Untersuchungen in einer
turbulenten Kanalstr\"omung
mit starken viskosen Wand\-schichten,''
Mitteilungen aus dem Max-Planck-Institut f\"ur Str\"omungsforschung und
der Aerodynamischen Versuch\-sanstalt, G\"ottingen, no. 48 (1970).\par

\ref{[3]}  H. Eckelmann,
``The structure of the viscous sublayer and the adjacent wall region
          in a turbulent channel flow,''
J. Fluid Mech., {\bf 65} (1974), 439--459.\par

\ref{[4]}  T. Wei and W.W. Willmarth,
``Reynolds-number effects on the structure of a
turbulent channel flow,'' J. Fluid Mech., {\bf
204} (1989), 57--95.

\ref{[5]}  R.A. Antonia, M. Teitel, J. Kim and L.W.B. Browne,
``Low-Reynolds-number effects in a fully developed
turbulent channel flow,'' J. Fluid Mech., {\bf 236} (1992), 579--605.

\ref{[6]}  J. Kim, P. Moin and R. Moser,
``Turbulence statistics in fully developed channel flow at low
          Reynolds number,'' J. Fluid Mech., {\bf 177} (1987), 133--166.

\ref{[7]}  M.V.  Zagarola,
{\it Mean-Flow Scaling of Turbulent Pipe Flow},
Ph.D thesis, Princeton University (1996).

\ref{[8]}  J.O. Hinze, {\it Turbulence}, Mc-Graw-Hill: New York, 2nd
edition (1975).

\ref{[9]}  R.L. Panton, 
``A Reynolds Stress Function for Wall Layers,''
J. of Fluids Eng., {\bf 119} (1997), 325--329.

\ref{[10]}  See, e.g. G.I. Barenblatt, A.J. Chorin and V.M.
Prostokishin,
``Scaling laws for fully developed turbulent flow in pipes,''
Appl. Mech. Rev., {\bf 50} (1997), 413, for a recent
survey of pipe flows and G.I. Barenblatt and A.J. Chorin,
``Scaling laws and vanishing viscosity limits
in turbulence theory,'' SIAM Rev.,
{\bf 40},No. 2, (1998), 265.

\ref{[11]} S. Chen, C. Foias, D.D. Holm, E. Olson, E.S. Titi and S.
Wynne,
``The Camassa-Holm equations as a closure
model for turbulent channel and pipe flow,'' Phys. Rev. Lett., {\bf
81} (1998), 5338.

\ref{[12]} W.V.R. Malkusa and L.M. Smith,
``Upper bounds on functions of the dissipation rate
in turbulent shear flow,''
J. Fluid Mech., vol. {\bf 208} (1989), 479--507.

\ref{[13]}  D. D. Holm, J. E. Marsden and T.S. Ratiu,
``The Euler-Poincar\'e Equations in Geophysical Fluid
Dynamics,'' in {\it Mathematics of Atmosphere and Ocean Dynamics},
Eds. M. Cullen, J. Norbury and I. Roulstone,
Cambridge University Press, 1999.

\ref{[14]}  D.D. Holm, J.E. Marsden and T.S. Ratiu,
``Euler-Poincar\'e Models of Ideal Fluids with Nonlinear
Dispersion,'' Phys. Rev. Lett.,
{\bf 80} (1998), 4173-4176.

\ref{[15]}   D.D. Holm, J.E. Marsden and T.S. Ratiu,
``Euler-Poincar\'e equations and semidirect
products with applications to continuum theories,''
Adv. in Math., {\bf 137} (1998), 1-81.

\ref{[16]}  R. Camassa and D.D. Holm,
``An integrable shallow water equation with
peaked solitons,''
Phys. Rev. Lett., {\bf 71} (1993), 1661--1664.

\ref{[17]} D.D. Holm, S. Kouranbaeva, J.E. Marsden,
T. Ratiu and S. Shkoller,
``A nonlinear analysis of the averaged Euler equations,''
Fields Inst. Comm., Arnold Vol. 2, Amer. Math. Soc., (Rhode
Island), (1999) to appear.

\ref{[18]}  S. Shkoller, ``Geometry and curvature of
  diffeomorphism groups with $H^1$ metric and mean hydrodynamics,''
  J. Func. Anal., {\bf160} (1998), 337-355.

\ref{[19]}  Y. Brenier, ``The least action principle and the
  related concept of generalized flows for incompressible perfect
  fluids,'' J. Amer. Math. Soc., {\bf 2} (1989) 225-255.

\ref{[20]}  A.I. Shnirelman, ``Generalized fluid flows, their
  approximation and applications,'' Geom. Func. Anal., {\bf 4}
  (1994) 586-620.

\ref{[21]}  J.E. Marsden, T. Ratiu and S. Shkoller,
``The geometry and analysis of the averaged Euler equations and a new
diffeomorphism group,'' Geom. Func. Anal., to appear.

\ref{[22]}  I. Gjaja and D.D. Holm,
``Self-consistent Hamiltonian dynamics of wave
      mean-flow interaction for a rotating stratified
      incompressible fluid,''
Physica D, {\bf 98} (1996), 343--378.

\ref{[23]}  J.E. Dunn and R.L. Fosdick,
``Thermodynamics, Stability, and Boundedness
    of Fluids of Complexity 2 and Fluids of Second Grade,''
Arch. Rat. Mech. Anal., {\bf
56} (1974), 191--252.

\ref{[24]}  J.E. Dunn and K.R. Rajagopal,
``Fluids of Differential Type:
    Critical Reviews and Thermodynamic Analysis,''
Int. J. Engng. Sci., {\bf
33} (1995), 689--729.

\ref{[25]}  R.S. Rivlin,
``The Relation Between the Flow of Non-Newtonian
    Fluids and Turbulent Newtonian Fluids,''
Q. Appl. Math., {\bf 15} (1957), 212--215.

\ref{[26]} A.J. Chorin,
``Spectrum, Dimension, and Polymer Analogies in
        Fluid Turbulence,''
Phys. Rev. Lett. {\bf 60} (1988), 1947--1949.

\ref{[27]}  T.H. Shih, J. Zhu and J.L. Lumley,
``A New Reynolds Stress Algebraic Equation Model,''
Comput. Methods Appl.
Mech. Engng., {\bf 125} (1995), 287--302.

\ref{[28]}  A. Yoshizawa,
``Statistical Analysis of the Derivation of the
        Reynolds Stress from its Eddy-viscosity Representation,''
Phys. Fluids, {\bf 27} (1984), 1377--1387.

\ref{[29]}  R. Rubinstein and J.M. Barton,
``Nonlinear Reynolds Stress Models and the
        Renormalization Group,''
Phys. Fluids A, {\bf
2} (1990), 1472--1476.

\ref{[30]}  A.A. Townsend, {\it The Structure of Turbulent Flows},
Cambridge University Press $\hfill\break$  (1967).

\ref{[31]}  J.L. Lumley and H. Tennekes,{\it A First Course in
Turbulence}, MIT Press (1972).

\ref{[32]}  M. Abramowitz and I.A. Stegun, {\it Handbook of
Mathematical Functions}, Dover Publications, Inc., New York, 9th
edition.

\ref{[33]} S. Chen, C. Foias, D. D. Holm, E. Olson
E.S. Titi and S. Wynne, ``The Camassa-Holm Equations and Turbulence,''
Physica D, 1999 (in press).

\ref{[34]}  L.D. Landau and E.M. Lifshitz, {\it Fluid Mechanics},
Pergamon Press, 2nd edition (1987).

\endbibliography

\vfill\eject

\noindent {\bf Figure Captions}

\bigskip

\noindent
FIG.~1.
The mean velocity profile
in the channel for the
constant-$\alpha $ viscous Camassa-Holm equation compared with the
experimental data of Wei and Willmarth.$^{4}$

\bigskip

\noindent
FIG.~2.
Reynolds shear stress
in the channel compared with the experimental data of Wei and
Willmarth.$^{4}$

\bigskip

\noindent
FIG.~3.
The statistical
compatibility of the VCHE with the theory of Panton$^{9}$ in the near
wall region.

\bigskip

\noindent
FIG.~4.
The mean velocity profile
in the pipe for the
constant-$\alpha $ viscous Camassa-Holm equation compared with the
experimental data of Zagarola.$^{7}$

\vfill\eject

\midinsert
\vskip.5\baselineskip
$$\vcenter{\hbox{\psfig{figure=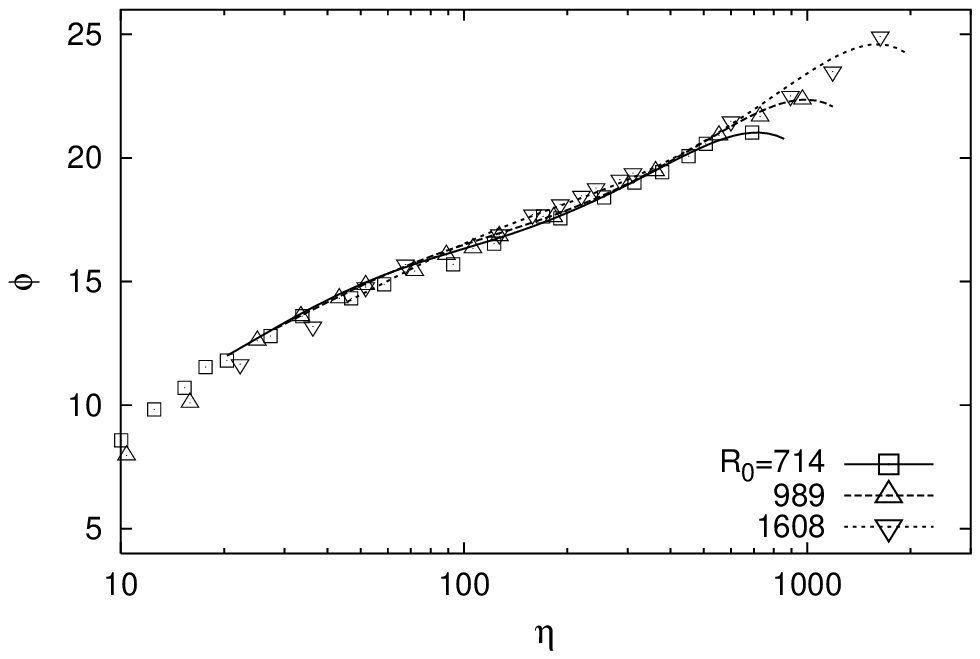,height=3truein}\hskip.85truein}}$$
\vskip.5truein
$$\vcenter{\box0}$$
\endinsert

\vfill\eject

\midinsert
\vskip.5\baselineskip
$$\vcenter{\hbox{\psfig{figure=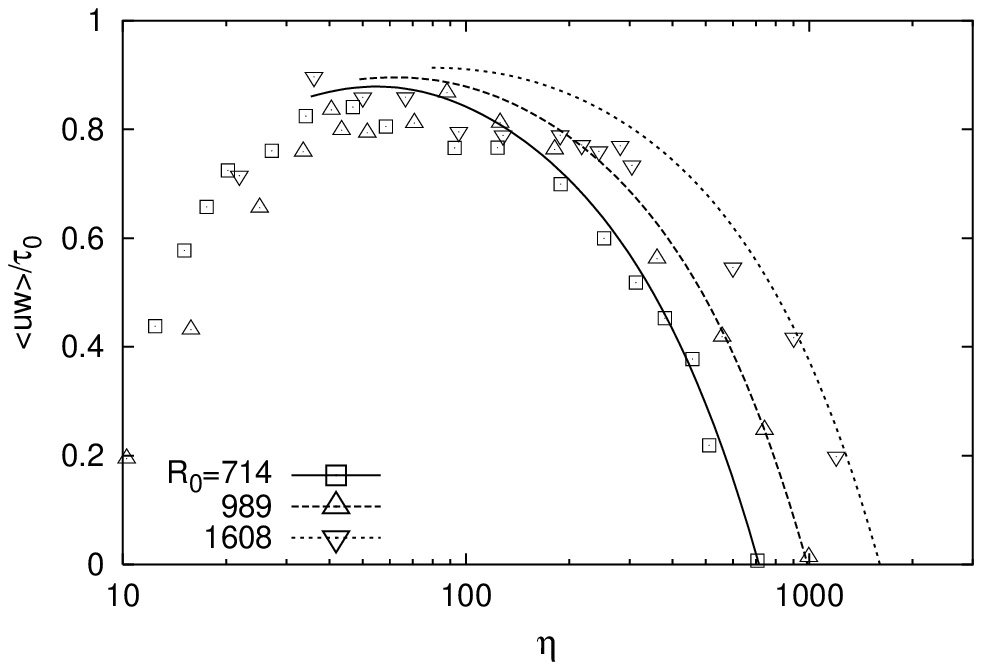,height=3truein}\hskip.85truein}}$$
\vskip.5truein
$$\vcenter{\box2}$$
\endinsert

\vfill\eject

\midinsert
\vskip.5\baselineskip
$$\vcenter{\hbox{\psfig{figure=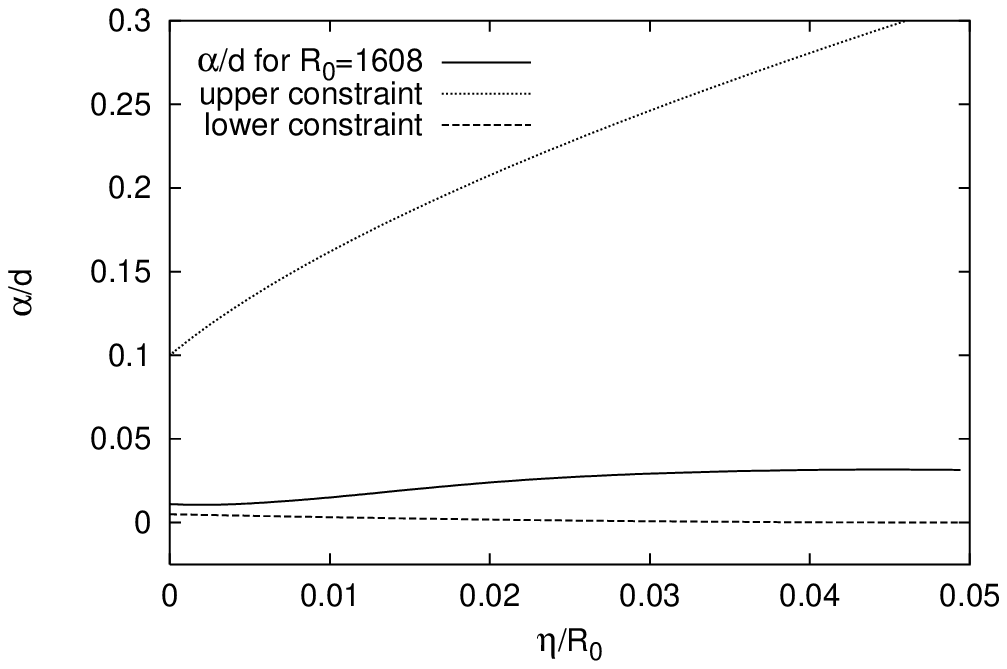,height=3truein}\hskip.85truein}}$$
\vskip.5truein
$$\vcenter{\box3}$$
\endinsert

\vfill\eject

\midinsert
\vskip.5\baselineskip
$$\vcenter{\hbox{\psfig{figure=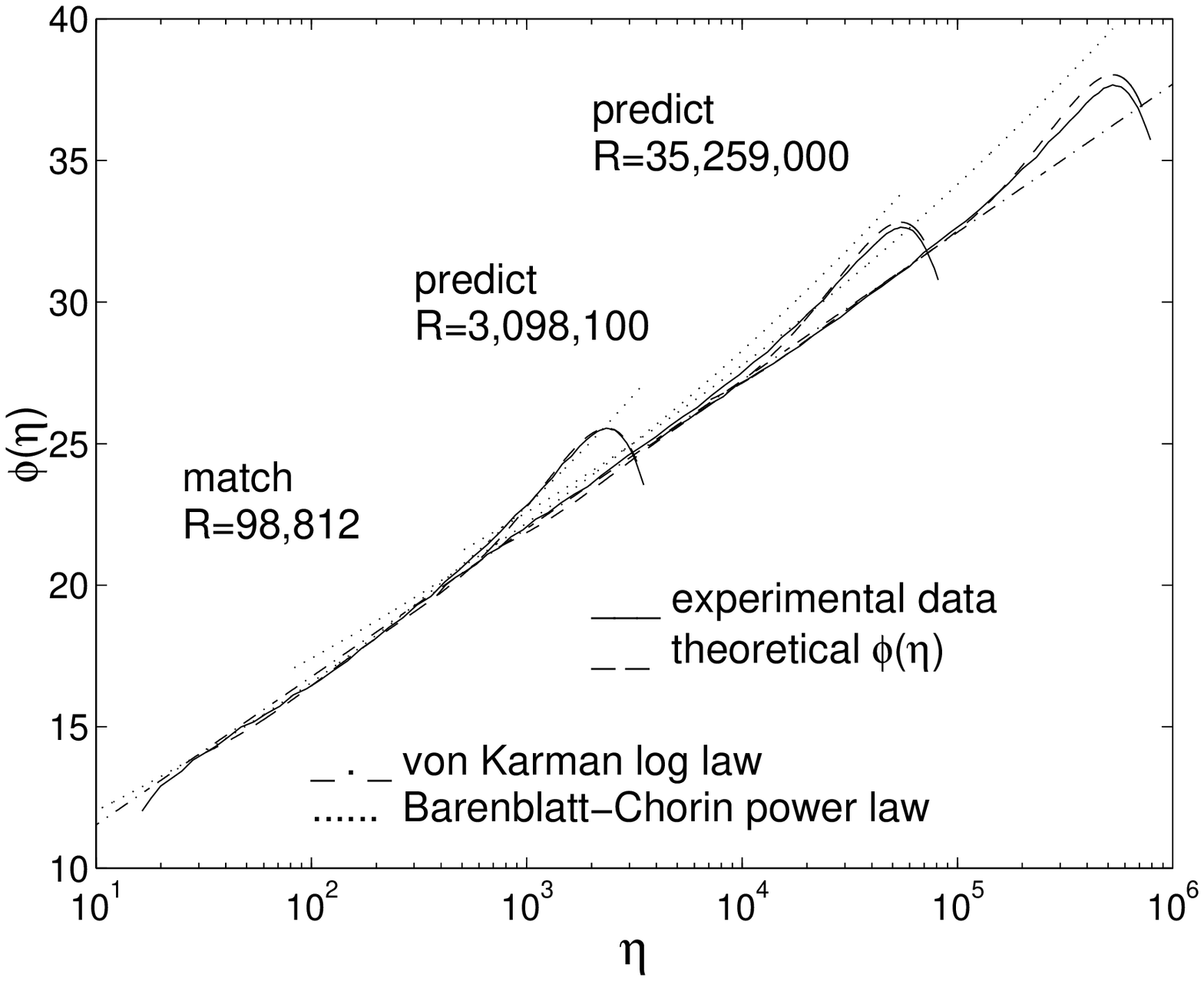,height=3truein}\hskip.85truein}}$$
\vskip.5truein
$$\vcenter{\box4}$$
\endinsert

\end